\documentclass[10pt]{article}
\usepackage[dvips]{graphicx}
\begin{document}

\newcommand{\m}[1]{\ensuremath\mbox{\boldmath $#1$}}
\newcommand{\be}{\begin{equation}} \newcommand{\ee}{\end{equation}}
\newcommand{\ba}{\begin{eqnarray}} \newcommand{\ea}{\end{eqnarray}}
\newcommand{\nn}{\nonumber} \renewcommand{\bf}{\textbf}
\newcommand{\ra}{\rightarrow} \renewcommand{\c}{\cdot}
\renewcommand{\d}{\mathrm{d}} \newcommand{\diag}{\mathrm{diag}}
\renewcommand{\dim}{\mathrm{dim}} \newcommand{\D}{\mathrm{D}}
\newcommand{\integer}{\mathrm{integer}}
\newcommand{\LL}{\mathbf{\Lambda}} \newcommand{\R}{\mathbf{R}}
\renewcommand{\t}{\mathrm{t}} \newcommand{\T}{\mathbf{T}}
\newcommand{\V}{\mathbf{V}} \newcommand{\tr}{\mathrm{tr}}
\newcommand{\cA}{\cal A} \newcommand{\cB}{\cal B}
\newcommand{\cC}{\cal C} \newcommand{\cD}{\mathrm{\cal D}}
\newcommand{\cF}{\cal F} \newcommand{\cG}{\cal G}
\newcommand{\cL}{\cal L} \newcommand{\cO}{\cal O}
\newcommand{\cT}{\cal T} \newcommand{\cU}{\cal U}
\newcommand{\s}{\,\,\,} \renewcommand{\a}{\alpha}
\renewcommand{\b}{\beta} \newcommand{\e}{\mathrm{e}}
\newcommand{\eps}{\epsilon} \newcommand{\f}{\phi}
\newcommand{\fr}{\frac} \newcommand{\g}{\gamma} \newcommand{\h}{\hat}
\renewcommand{\i}{\mathrm{i}} \newcommand{\p}{\partial}
\newcommand{\w}{\wedge} \newcommand{\x}{\xi}
\def\slash#1{\setbox0=\hbox{$#1$}#1\hskip-\wd0\hbox to\wd0{\hss\sl/\/\hss}}

\input{epsf}

\centerline{\large\bf {Exploring the Micro-Structure of the Proton:}} 

\centerline{\large\bf {from Form Factors to DVCS}}

\bigskip

\centerline{\large \bf {John P. Ralston$^a$ and Pankaj Jain$^b$}}

\begin{center}
$^a$ Department of Physics and Astronomy,\\ University of Kansas, Lawerence,
KS 66045, USA\\
$^b$ Physics Department, I.I.T. Kanpur, India 208016
\end{center}

{\bf {Abstract:}} For a long time people made the mistake of thinking
the proton was understood. New experiments, ranging from form factors
to deeply virtual Compton scattering, promise a new era of highly
informative studies.  Among the controversial topics of the future
may be such basic features as the physical size of the proton, the
role of quark orbital angular momentum, and the possibility of making
"femto-photographic" images of hadronic micro-structure.

\section*{Reflections on the First Form Factor}

{\it Apology:} Hadronic physics is still something young.  And yet,
people thought they understood the proton for a long time.  This was
not right, but persisted because so little was known.  When little is
known, we cannot even find out what {\it might be known.}

Now we face a new time, an era promising informative measurements
on how hadrons are made.  We should stand back, and assess how
hadronic physics came {\it not to be understood} up to this point.  I
apologize in advance for needing to explain things gone awry at an
elementary level.  I will review some history, from ancient to current
day, to set the stage for new developments exploring the
three-dimensional micro-structure of hadrons.

Rethinking the thinking about form factors led to the question: what
was the first form factor?  Newton gave us the
gravitational form factor of the Earth.  College students should
repeat the integration exercise, assuming a uniform density.  You
cannot cheat and use Gauss' Law, because Gauss was not yet born.

The claims about the form factor were probably met with some
skepticism by the gentlemen of the Royal Society.  First, the form
factor describes an incredibly unnatural theory with an exceedingly
small and arbitrary parameter.  The theory explained little new in
terms of phenomena: things falling to the Earth being already known.
The main parameter was made absurdly small to escape from direct
observation of the claimed universal force.  Then there had to be
absurdly large parameters, such as the mass of the Sun, to compensate
the small parameter.  The theory's author sidestepped direct tests,
and based results on astronomical data\ldots and we all know how
unreliable that kind of data can be!  Following indirect arguments,
and inventing private mathematical methods to justify it, Newton
claimed that the entire Earth form factor could just be considered
``the same as a point mass at the center'', {\it a perfectly
incredible result}.

While the form factor acted like a point mass, nobody of course
believed that the Earth was a point mass.

Cavendish was a hard-minded, bitter experimentalist who could go out
and make the measurements that Newton shirked.  The ``Honorabilus
Henricus Cavendish'' enrolled at Peterhouse College of Cambridge at
age 18 in 1749.  He spent years on the Form Factor problem.  Henry
needed the Form Factor in the measurement of gravity from mountains.
He chose a special hill, ``Schieshallion'', because of its near
conical form and calculability.  The story that Cavendish ``weighed
the Earth'' is only partly correct.  Cavendish must have had a pretty
good idea of the size of Newton's $G$ from the start.  Just equate
$g=9.8 m/s^2$ with $GM_E/R^2 $.  The Earth's mass $M_E$ is estimated
to be $4 \pi R^3 \rho /3$, where $\rho$ is the local density of rocks,
something like 2-3 times the density of water.  So Henry had $G \sim
3g/(4\pi R\rho)$ before he ever started his experiment.  When Henry
got an answer for $G$ several times larger, he did not believe it was
anything fundamental.  In experimental tradition, he {\it complicated
the form factor rather than challenge the theory}.  Now we say to ``go
to larger momentum transfer where there will be a dense core''.  More
than 200 years later this speculation remains untested by direct
means.  Seismology is very indirect, and what lies inside the Earth
remains mysterious, although we may someday look inside with neutrino
tomography \cite{tomo}.

The modern era produced the form factors of hadrons.  People say that
Rutherford postulated the central nucleus by looking at his data.
Reading the actual paper \cite{Rutherford} is very enlightening.  One
learns there was already an excuse for large scattering angles, coming
from J. J. Thompson invoking multiple scattering in the plumb-pudding
model.  Rutherford's paper is phenomenological, referring to data of
Geiger and Marsden: he demolishes the proposal of multiple scattering
on statistical grounds \cite{Rutherford}.  When Rutherford proposes
the nuclear center, he was perhaps not the first, citing ``\ldots it
is of interest that Nagoaka (Phil Mag vii, 441(1904)) has
mathematically considered a `Saturnian atom' (with ) \ldots rings of
electrons''.  Predictably, there is a damaging typographic error, and
it is right in the famous Rutherford formula for scattering!

Anyway, by 1929 the finite size of the strongly interacting
nucleus was known.  At the Royal Society in London,
Rutherford stated \cite{RuthQuote} ``It will be seen that this (data) makes the
nucleus of Uranium very small, about $7 \times 10^{-13} cm$\ldots it
sounds incredible but it may not be impossible''.  Of course the
hyperfine splittings of atomic physics gave similar numbers, and
Yukawa knew it to predict his mesons.

\medskip

In other words, everybody knew that the proton had a finite size for a
long time. The ``size'' was hardly open to much question.

\subsection{Recent Prehistory: Electron Beams}

Our era is dominated by Robert Hofstadter's spectacular measurements
of the proton form factors via electron scattering.  The two form
factors are called $F_1$, $F_2$ and defined by \ba <p', \, s'| J^{\mu}
|p,\, s> = \bar u(p', \, s') \ [ \gamma^{\mu} F_1(Q^2)+
\frac{\kappa}{2M} F_2(Q^2)\sigma^{\mu \nu}Q_{\nu} \ ] u(p, \, s) .
\ea Here $Q^{2}$ is the momentum transfer-squared; it is spacelike
(negative) in electron scattering.  One famous Hofstadter electron
beam used the heroic energy of 188 MeV. Nobody faulted Hofstadter's
data \cite{Hofstadter}, as far as I know, and the Nobel Prize of 1961
seems perfectly appropriate.  But Robert already had the charge radius and its
interpretation before he started his experiment.

Meanwhile there were serious questions about this specific {\it
interpretation } of the form factors.  I am summarizing all
this history just to bring out these {\it interpretation problems
which persist today.} The question is whether the low $Q$ form factors
measure the Fourier transforms of charge and magnetic moment
distributions.  As far as I can tell, people with savvy backed away
from this in the old days, yet many believe it today.  The
interpretation might be justified: I just don't find credible
justification for what has become an important point.

   The influential book of Drell and Zachariasen \cite{Drell} said {\it
   ``it is convenient to define''} densities by Fourier transform in the
   Breit frame, following this by {\it ``\ldots this definition depends
   on a particular definition of Lorentz frame in which the proton is
   not stationary, and therefore the relation of these densities to any
   real physical extent of the proton is quite unclear''.} (It is also
   puzzling that the Breit frame $Q^{\mu}=(0, \, 0,\,0,\, Q_3)$ is very
   far from unique, and you will get the very same photon momentum with
   two sideways Lorentz boosts of arbitrary rapidity.)  The target's
   ``mean-square radius'' was stated to be ``just measuring the slope of
   the form factor as $Q^2\ra 0$.''  This is a definition and
   meaningless.  Sakurai's book \cite{Sakurai} shows that the form
   factors {\it could be reproduced} by the claimed static
   distributions, without saying that such static distributions had been
   {\it measured}.  Other books (e.g. Schweber \cite{Schweber}) are very terse,
   avoiding interpretation.

Apparently people became divided, between those accepting the form
factors were Fourier tranforms of real spatial distributions, and
those who found this misleading.  An important review article by
Yennie, Levy and Ravenhall \cite{Yennie} was unequivocal, stating:

   \medskip

   {\it Objection 1:}{\bf { ``The essential point is that measurement of
   structure within a distance $d$ requires values of $|\vec Q|$ of
   order $1/d$\ldots} }

and going on to say:

   \medskip

{\it Objection 2:}{\bf {\ldots and if absorption of this momentum
causes recoil (i.e if $|\vec Q| > Mc/\hbar$) the intuitive concepts of
static charge and current distributions are no longer valid.''}}

\medskip

The labels of Objection 1 and Objection 2 are my own.  Yennie doubly
rejected the identification of form-factor as charge density.  Indeed
the Appendix of Yennie's important 1959 {\it Reviews of Modern
Physics} \cite{Yennie} lists two other form factors denoted $A$ and 
$B$, now called
$G_M$ and $G_E$, which are linear combinations of Rosenbluth's: \ba
G_M = F_{1}+ \kappa F_{2};  \\ G_E =F_{1}+
\kappa \tau F_{2}; \:\: \tau=\frac{Q^{2}}{4 m_{p}^{2}}.\label{sachs}\ea
(Sachs cites this when he writes his own article: perhaps Yennie {\it
et al} found the Sachs form factors first.)  Yennie {\it et al} gave
the alternative linear combinations just to emphasize that the
connection of form factor and charge density was highly arbitrary.  A
few years later \cite{Stanford1963}, Yennie reiterated his stance,
adding that ``\ldots Sachs and collaborators do hold there is a real
physical meaning'' to the Fourier transform relation.  Reading Sachs'
paper \cite{Sachs} gives little clue why Sachs was so vehement in the
claim, or why people believed it.

Objection 2 is not a worry if the momentum transfer is small compared
to the mass.  This seems to have diverted attention.  Meanwhile
Objection 1, which is the objection of {\it optical resolution} and
indeed the entire problem of {\it dynamics}, is swept under the rug!

Right on the heels of the Hofstadter data, dynamics forced the
interpretation in terms of VMD, {\it vector meson dominance}.  The
isoscalar and isovector mesons were predicted \cite{Frazer} by 
saturating the photon
exchange with mesons.  Physically, VMD means that the meson {\it
strongly polarizes} the proton by resonant pion exchanges, which then
re-interact by resonant pion exchanges.  What scrap of the proton left
over \ldots that is not the meson\ldots is not even part of the model!

How then is the ``charge radius'' related to the {\it undisturbed
radius of the proton}?  Throwing pillows at a baseball, do you measure
the size of the baseball or the size of the pillows?  The ``most
general possible'' nature of the Rosenbluth formula subconsciously
convinced many that the question did not matter.  Hyperfine and Lamb
shift experiments use the same matrix element, so there is nothing but
trivial consistency there.\footnote{If, however, the dressing of the
electron in the bound state by non-perturbative effects is
substantially different from that in the scattering state, then the
two measures might disagree.  This is one reason that precision low
$Q$ measurements remain exceedingly important.}

Yet Objection 1 was really about {\it optics}.  Following Yennie, it is
hard to believe that a momentum transfer of 200 MeV could give
anything else but a distance scale of 1 Fm.  The measurement of 200
MeV put in a Fermi scale and came out with a Fermi scale.  In what
sense was the size of the proton measured?

Since the process is dynamical, I arrived at an idea that perhaps some
of the ``pion cloud'' is actually stirred up by the strong interaction
itself, and was not there in the proton to start with.  In fact,
dynamical strong polarization effects do occur for atoms on metal
surfaces, where measured dipole moments per charge can be {\it
thousands of times larger} than the physical size of the atoms.  It is
explained by fast electrons free to move quickly on the surfaces.  In
the 1960's nobody had a clue that there existed light, relativistic
quarks, but now we know these quarks exist, and they react on time
scales much faster than an inverse 200 MeV.

Yet the idea of an intrinsic 1 Fm pion cloud around nucleons is deeply
ingrained.  It takes effort and initiative to question what we may or
may not know about it.  Everyone knows that a static system coupled to
a meson of mass $\mu$ uses a Green function $1/(\vec k^{2}+\mu^{2})$
and cannot fall off with distance {\it less} rapidly than $e^{-\mu
r}$.  This says nothing about {\it all the rest} of the interior and
dynamical structure, for which optical resolution much better than
$1/\mu$ is required to say {\it anything}.

Imagine how interesting it would be if the undisturbed size--the {\it
quark radius} of the proton -- is not the same as the so-called charge
radius measured via VMD. This would be a gold mine, not only for what
the proton {\it is}.  It would be a gold-mine of classic scientific
controls.  When distinct things can be compared, the standard
measurements called the charge radii become that much more valuable
and interesting.  And so the beautiful experimental accomplishments of
the neutron charge radius \cite{neutron}, for instance, would not just
be isolated marvels but would be something to compare and to teach us
more.  One needs large $Q^{2}$, of course, to pursue this.

We then move to the modern description of large $Q^{2}$ form factors.
Unfortunately the topic has little to do with the bulk of the Fock
space components making up the proton.  However we will return to
probing the overall makeup of the proton afterwards.

\section{Hadronic Form Factors at Large $Q^{2}$}

Our approach to large $Q^{2}$ form factors in pQCD uses impact
parameter factorization \cite{impact,PR90}.  In this scheme the
transverse spatial separation $\vec b$ between quarks, as well as the
longitudinal momentum fraction $x$, is used to describe amplitudes.
The impulse approximation, which we consider the hallmark of pQCD, is
used to separate a hard-scattering kernel from wave functions (or more
general correlations) with longer time-scales.  In the impulse
approximation the $k^{-}$ integrations are done to evaluate products
at light-cone time zero.  The expression for a baryon form-factor $F$
scattering 3 quarks takes the form \ba F(Q^{2}) = \int
d^{2}b_{1}d^{2}b_{2}dx_{1}\ldots dx_{4}\bar \psi(x_{i}, \, \vec b_{i};
\, Q)H(Q, \, x_{i}, \, \vec b_{i}) \psi(x_{i}, \, \vec b_{i};\, Q).
\label{factored} \ea Note the following:

\medskip

$\bullet$ This method does not make prior assumptions about short
distance.  Current research in pQCD has found that the assumption that
all $b_{i}\ra 0$ is neither justified nor necessary.

Theorists started with $b\ra 0$.  However the old ``operator product
expansion" (OPE) was shown unable to account for numerous physically
observable effects, such as the Sudakov corrections \cite{LiSterman,Li}.
In Eq. \ref{factored} the Sudakov effects are merged into definitions
of the wave functions: elsewhere \cite{LiSterman,Li,ourExclusives} they
are denoted $e^{-S(x_{i}, \, \vec b_{i};\, Q)}.$

$\bullet$ The method explains the constantly observed phenomena
\cite{HelFlip} known as hadronic helicity flip, in which the sums of
the helicities of the hadrons going into the reaction does not equal
the sum of those going out.  For a while people had the impression
helicity flip was not part of pQCD: it is, in fact, a mainstream part,
and quantitative calculations of hadronic helicity flip in pQCD are in
the literature \cite{Gousset}.  The cause for misunderstanding is
over-reliance on $b\ra 0$ in quick asymptotic estimates.

$\bullet$ Quark-counting must play a role, as originally envisioned by
Brodsky and Farrar and Matveev {\it et al} \cite{BFMatveev}.  Under
this hypothesis, and as confirmed by perturbation theory, the addition
of extra quanta beyond the valence {\it contains important regions}
that produce suppression at large $Q$.  The topic has been
controversial, and ```strongly polarized'', but the power-law scaling
observed in so much data is not to be dismissed: the data exists
despite many theoretical complaints \cite{ILLS}.

Yet we avoid the later {\it asymptotic short distance} (ASD) approach
of Brodsky and Lepage \cite{BL} sometimes said to be the same thing as
QCD. It is not the same thing.  The ASD formalism is decisively ruled
out by observing hadronic helicity flip\ldots and ruled out many
times.  In particular, $F_{2}=0$ in the ASD formalism for massless
quarks in valence state.  Hadron helicity violation is too universal
and $F_{2}$ is too large to be explained by flipping the spins with
perturbative quark masses $\sim$ few  MeV: so drop ASD.

$\bullet$ The impact-parameter coordinate has been around since the
beginning of time.  Our usage was influenced by the Sterman and Botts
\cite{impact} treatment of the Landshoff independent scattering
processes of $pp$ elastic scattering.  Independent scattering is yet
another example of something beyond the capacity of the ASD formalism.
Impact-parameter factorization also allows the systematic pQCD
description of color transparency \cite{PR90,JainR93} and nuclear
filtering \cite{physRep} which are beyond ASD. Subsequent to our
treatment of {\it nuclear} form factor transparency \cite{PR90}, it
was realized that {\it free space} hadron form factors needed the
impact parameter factorization, too \cite{LiSterman,Li}.

By similar care, the helicity flip terms are also saved for evaluation
in pQCD \cite{Gousset}.  Helicity flip calculations are ``leading
twist'' for what they describe.  It is grossly {\it misleading} to
compare them to ``higher twist'' meaning the subleading corrections to
other matrix elements with different transformation properties.
This is why the approach of Eq. \ref {factored} constitutes a
different and more general theory of hadron scattering than the
asymptotic short distance one.  By now most groups studying
quark-counting use the impact parameter factorization.

\subsection{Quark Orbital Angular Momentum}

A compelling motivation for using impact parameter coordinates is that
it allows classification of wave functions in orbital angular
momentum.

The quantization $z$-axis is along the direction of particle momentum.
The natural orbital angular momentum of high energy processes uses the
Lorentz-invariant subgroup SO(2) of rotations about the z-axis.  We do
not discuss SO(3) representations, which do not transform well.  The
representations of SO(2) orbital angular momentum (OAM) are labeled by
an integer $m$, the eigenvalue of an operator $L_{z}.$ The expansion
of wave functions in OAM is conventionally written $$\psi(x_{i}, \,
\vec b_{i};\, Q)=\sum_{m}e^{i m\phi}\psi_{m}(x_{i}, \, |\vec
b_{i}|;\,Q).  $$ On general grounds of continuity, the expansion
coefficients $\psi_{m}(x_{i}, \, |\vec b_{i}|;\,Q)$ obey a power rule,
\ba \lim_{b\ra 0}\, \psi_{m}(x_{i}, \, |\vec b_{i}|;\,Q) \sim b^{|m|}.
\ea We call this the ``$b^{m}$ rule''.  If a model takes the limit
$b\ra 0$ in the first step, only $m=0$ survives, and all information
about quark OAM is forever lost.  Historically this faulty limiting
procedure \cite{BL} underlies the misconception that hadron helicity
flip could not be treated straightforwardly in pQCD.

Despite popular belief, there is {\it no power-counting
preference} for high-energy wave functions to be in the $m=0$
``s-wave''.  For a simple example, consider the $q \bar q$
Bethe-Salpeter wave function $\psi_{\pi}$ for a pion of momentum $p$. 
There are four invariant
wave functions: 
\ba \psi_{\pi} = A \slash{p}\gamma_{5}+ B [\slash{p},\,\slash{b}]\gamma_{5}+
C \slash{b}\gamma_{5} + D\gamma_{5} \label{wavefuns}\ea 
The first two terms
(A, B) scale equally like $p$ in the high energy limit, representing
the ``large'' components of the quark operators.  Indeed scaling like
$p$ is the largest possible result one can get from two Dirac spinor
field operators.  Note that $B$ carrys $L_{z}=\pm 1$, as seen by
expanding $\vec b_{T}=b_{x}\pm i b_{y}$.  So orbital angular momentum
shows up in the {\it transverse plane}.  By angular momentum counting,
$L_{z}=\pm1$ is the maximum for the pion ($| L_{z}| <2 $ for the
proton).  The explicit factors of $b$ in Eq. \ref {wavefuns} obey the
$b^{m}$ rule, and the series expansions of coefficient functions $A,
\, B, \,\ldots$ start at $b^{0}$ unless a model has extra selection
rules.  There are also small terms, namely $C$ and $D$, suppressed by
$1/p$ in comparison.  Vector mesons and baryons are described
similarly: it is a nice exercise to find the eight (8) covariant wave
functions of a vector meson \cite{Gousset}.

How can one test directly if mesons contain quark OAM? We suggest
nuclear filtering for $A>>1$ in the reactions of $\gamma^{*}(Q)+ A \ra
meson+ A^{*}$ \cite{Dourdan}.  The meson should carry all of the
virtual photon energy $\nu$ up to resolution.  The final nucleus
$A^{*}$ can be disrupted, and the reaction need only be exclusive to
the extent no extra pions escape.  For $Q^{2}> {\rm GeV}^{2}$ this
reaction involves sizable momentum transfer to the struck nucleon.
The ``big fat'' $m \neq 0$ wave functions should be relatively
depleted by filtering \cite{physRep} compared to lean and mean $m=0$
types.  To be specific, the leading short-distance vector mesons are
longitudinal, so that the ratio $\sigma_{L}/\sigma_{T}$ should
increase with increasing $A$ and fixed $Q^{2}, \,\nu$.  Everything we
have learned from color transparency \cite{physRep} says this
filtering effect will be much more dramatic than the corresponding
transparency effect of increased $\sigma_{L}/\sigma_{T}$ with fixed
$A$ and increasing $Q^{2}, \,\nu$.  Indeed, small $A \sim 10$ is not
nearly as good as large $A$, because small $A$ is dominated by edge
effects.

Why do we not get the OAM wave functions by making a Lorentz boost?
The boost operator in field theory is exponentially complicated in the
Poincare generators: it creates interacting particles.  The fact
remains that our conventions are Lorentz invariant, and Gell-Mann's
quark model stands to be ruled out if quarks in the boosted proton
have a lot of orbital angular momentum.

\subsection{$F_{2}$ and Quark Orbital Angular Momentum}

The most interesting example of usage is surely $F_{2}$.  After seeing
the data of Jones {\it et al} \cite{jones}, which seemed to indicate
previous concept errors in the treatment of $F_{2}$, we returned to
predict that the ratio $QF_{2}/F_{1}$ should be constant \cite{Buniy,Bologna}.
The
subsequent measurements of Gayou{\it et al} \cite{Gayou} found a ratio
$QF_{2}/F_{1}$ spectacularly constant, sparking enormous new interest
in the subject.  It appears that $QF_{2}/F_{1}$ is a scientifically
pivotal quantity that will be important for years to come.  We now
explain our predictions and recent work.

Counting angular momentum as helicity in the high energy limit, the
{\it chirality flip} of $F_{2}$ also causes one unit of {\it physical
angular momentum flip} with corrections of order $m_{p}/Q$, where
$m_{p}$ is the proton mass.  Angular momentum conservation is
maintained by the virtual photon breaking the symmetry in a frame
where $Q^{\mu}=(0, \, \vec Q_{T}=\sqrt{-Q^{2}}, \, 0)$.  Now it is
inconsistent with the chiral symmetry of pQCD to allow a light quark
(mass $m_{q}$) to flip helicity (with corrections of order $m_{q}/Q= 5
{\rm MeV}/Q$).  It follows that there is a general $\delta L_{z}= \pm
1$ rule for the net quark {\it orbital} angular momentum change in the
process.  For very general reasons, then, we claim that {\it
$QF_{2}/F_{1} $ measures the strength of quark orbital angular
momentum.} \cite{Buniy,Bologna}.  We believe this will be dominated by
the $m=0, \, m=\pm1$ OAM interference terms.

The facts of OAM of course tie directly to the proton's ``spin
crisis'', which is replayed with every hadronic helicity flip deja vu
all over again.\footnote{Yogi Berra is acknowledged here.}

\subsubsection*{$F_{2}$ and the Moments of Hard Scattering Kernels}

Elsewhere we discussed $F_{2}$ in several ways: first from an early,
independent invention of generalized parton distributions \cite{BNL93}
(apologies to Dittes {\it et al} \cite{Dittes} of whom we were
unaware), then from wave function \cite{Buniy} and updated generalized
parton distribution (GPD) points of view \cite{Bologna}.  In the
current work \cite{inProgress}, we estimate $F_{2}$ using the kernel
and wave-function re-arrangements of Sterman and Li \cite{LiSterman}.
The kernels are the 48 full, complicated Feynman diagrams evaluated to
leading order in Fermion transverse momentum.  After transforming to
$b$ space (two integrals) there are four $x$ integrations left.  We
use the COZ distribution amplitudes \cite{COZ} as a model for the $x$
dependence, and the full Sudakov kernel; parameters and formulas are
given in Ref. \cite{ourExclusives,Li}.  Examination of the Dirac
algebra shows no selection rules preventing the claimed interference.
We estimated $QF_{2}/F_{1}$ by calculating the moments of $b$
inside the integrands, because it is particularly important to see if
$b \ra 0$ is or is not an outcome of the calculation.  The moments are
defined by \ba &&\frac{QF_{2}}{F_{1}} \sim \frac{<b^{m}>}{<b^{0}>};
\nn \\
&&<b^{m}>= \int d^{2}b_{1}d^{2}b_{2} \:\: b^{m }\:\: dx_{1}\ldots
dx_{4}\bar \psi(x_{i}, \, \vec b_{i} \, Q)H(Q, \, x_{i}, \, \vec
b_{i}) \psi(x_{i}, \, \vec b_{i};\, Q). \nn \\
\label{moments} \ea In this intimidating expression we have tried to
leave extra space to show that the ratio is simply the moment of
$b^{m}$ divided by the moment of $1=b^{0}$.

We use the same hard scattering kernels discussed for years for
$F_{1}$ and keeping all the diagrams of pQCD. By using Eq. \ref
{moments} we have assumed that the $x$ dependence of non-zero OAM is
close enough to standard models to make a reasonable
estimate.\footnote{None of the $x$ dependences of any exclusive light cone wave
functions are known anyway.  The $x$ dependence of deeply inelastic
scattering sums over all Fock components.} The calculation tests
whether there is any {\it kinematic and power-related} suppression of
OAM at laboratory values of $Q^{2}.$ The kernels $H$ contain two
factors of $K(x_{i}x_{j}\sqrt{Q^{2}b_{k}^{2}})$, modified Bessel
functions, and are given in the literature \cite{ourExclusives,Li}. The
previous thinking in
the field was that if $Q^{2}\ra \infty$ in the first step, we would
get $<b^{m}>\sim 1/Q^{m}$.  That step would be an asymptotic
estimate, exactly of the kind subject to the ASD limit interchange
problems discussed earlier, and we do not do it.  Instead we {\it
first calculate the integrals and then look at the limit}.

\begin{figure}
\includegraphics[scale=0.8]{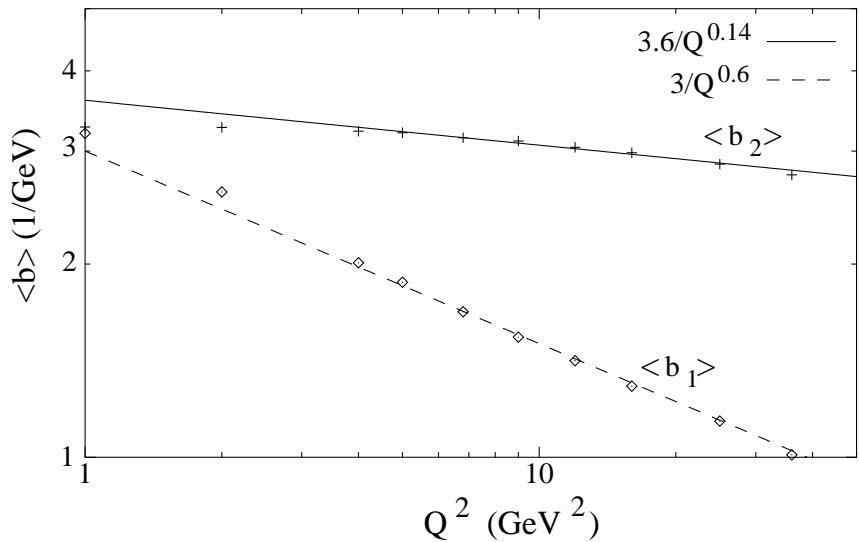}
\caption{The moments $<b_1>$ (dashed line) and $<b_2>$ (solid line),
as defined in the text, using the COZ distribution amplitude. We
see that the moment $<b_2>$ is almost flat in contradiction to the
ASD expectations.}
\label{fig:bmoments}
\end{figure}

The result is shown in Fig.  \ref{fig:bmoments}.  The figure shows
that $QF_{2}/F_{1}$ is supposed to be flat.  Certainly one moment is
nearly flat across the range of $Q^{2}$ accessible to existing and
future experiments.  We use the notation of Ref.  \cite{Li}.  The two
$up$ quarks are labelled as quark 1 and 2 and the $down$ quark is
labelled as quark 3.  The coordinate system is chosen such that the
quark 3 lies at the origin.  Let ${\vec B_1}$, ${\vec B_2}$, ${\vec
B_3}$ be the transverse positions of quark 1, 2 and 3 respectively.
Then we define the transverse separations $b_1$, $ b_2$ and $ b_3$ by
the relations, $ b_1= |{\vec B_1} - {\vec B_3}|= |{\vec B_1}|$, $ b_2=
|{\vec B_2} - {\vec B_3}|=|{\vec B_2}|$ and $ b_3= |{\vec B_1} - {\vec
B_2}|$.  In Fig.  \ref{fig:bmoments} we have plotted the moment of the
transverse separations $ b_1$ and $ b_2$.  We see that the moment $<
b_2>$ is almost flat.  The moment $< b_{1}>$ scales differently: this
is perfectly reasonable, because the 3 quark integration has regions
that are not symmetric.  Since our approach is very general, we
believe that the experiments finding $QF_{2}/F_{1}$ flat are seeing
quark orbital angular momentum in a very definitive way.  From the
normalization of $QF_{2}/F_{1}$, the $m \neq 0$ wave functions must be
normalized to a substantial fraction of the proton's spin, in accord
with our earlier observations \cite{BNL93,Bologna} and other work
\cite{KrollGPD}.

What does the constituent quark model say in this regard?  We refer
readers to a growing literature.  Frank and Miller \cite{MillerFrank1}
also predicted $QF_{2}/F_{1} \sim constant$ in 1994 on the basis of an
earlier model of Schlupf \cite{Schlupf}.  The prediction has recently
been updated and re-examined \cite{MillerFrank2}.
This model uses quark masses of about $260$  MeV, and focuses on
care in respecting relativistic spin-projections.  The
spin-projections are the key because in the end the model predicts
non-zero OAM on the light-cone, in remarkable concord with our much
different starting point.  Weber and collaborators \cite{weber} communicated
results of much the same kind at this meeting.

GPD's are also a good approach to the form factors, and a way to use
both constituent quark-model and short-distance concepts.  Afanasev's
\cite{Afanasev} pretty calculation inspired Stoler \cite{Stoler}, who
has shown that the data for $F_{1}, \, F_{2}$ can be fit in the GPD
formalism up to 30 GeV$^{2}$.  The starting point of this model does
not show the relation to OAM, but it is contained in the references.
Stoler shows that the data at the higher $Q^{2}$ regime requires a
hard wave function in GPD terminology, or a hard scattering kernel in
our language, or some original quark-counting (not ASD) contribution in
any language.  The results puts to rest claims \cite{ILLS,Rady} that
form factors are totally dominated by soft effects.  In regard to GPD,
we reiterate that there exists an infinite number of factorization
methods, of which the GPD are one example: all are capable of
representing pQCD, and it is not helpful for any one method or the
other to claim to be ``the unique'' pQCD prediction.

How can one test directly if the mysterious ratio $QF_{2}/F_{1}\sim
const$ is due to quark OAM? Use polarization transfer in heavy nuclei,
$\vec e +A \ra \vec p +A^{*}-1.$ It is a beautiful observable and an
experiment on Oxygen has already shown it to be feasible at several
GeV \cite{oxygen}.  Again one needs nuclear filtering with nuclear number
$A>>1$ to make any headway.  $A>>1$ means Gold, not Carbon, and the
door is open here.  If not keep hammering.  Filtering should deplete
the ``fatter'' $m\neq 0$ components of the wave function relative to
the $m=0$ component.  The polarization ratio corresponding to
$G_{E}/G_{M}$ should dramatically approach the ASD predictions for
$A>>1$.  In other words, $QF_{2}/F_{1}$ will drastically decrease with
large $A$ and large $Q^{2}$: the rate of decrease can tell us
something about the orbital configurations, with $|m|=2$ producing a
faster decrease than $|m|=1$.

\section{Beyond Form Factors: The $DVCS$ Microscope}

The interpretation of form factors at low $Q^{2}$ being soft,
theorists long requested large momentum transfer.  Yet large $Q^{2}$
form factors, however fascinating, have very little to do with the
undisturbed proton.  In any Lorentz frame, and in any model, the
proton is violently accelerated, and only a very tiny fraction of wave
function is involved in the final outcome.

However there happens to be {\it two kinds} of momentum transfer in
reactions with a virtual photon.\footnote{This section describes
recent work with Bernard Pire.}  Let $Q^{2}$ consistently denote the
{\it virtuality} of the photon, its 4-momentum-squared.  Let
$t=(p-p')^{2}$ denote the {\it momentum transfer to the target}.  In
form factors the kinematics is restricted to $Q^{2}=t$.

Meanwhile $Q^{2}\neq t$ in many reactions assuming that the {\it
momentum sent in by the virtual photon escapes} by some other path.
The most beautiful case is deeply virtual Compton scattering, (DVCS)
describing $\gamma^{*}+ p \ra \gamma _{p'}$, where $p, p'$ denote
hadron momenta.  Up to a kinematic $t_{min}$ restriction (and
$t_{min}$ goes close to zero for multi-GeV energies), reactions can have
$|Q^{2}| >{\rm GeV}^{2}$ with $t_{min} \sim 0 < |t|< {\rm GeV}^{2}$ 
in the
laboratory.  Much the same kinematics apply to virtual meson
production, in which $\gamma^{*}$ is effectively converted to a meson
such as a pion, rho or phi.

DVCS has caused a great deal of excitement because these reactions
involve GPD's.  Moreover, GPD's are diagonal in good coordinate bases,
namely the impact parameter representation, allowing a probability
interpretation \cite{Burkardt,Buniy,Diehl}.  However most studies had to
assume that a GPD model would be (1) guessed by theory, (2) integrated
with the quark and gluon kernels, (3) weighted by the quark and gluon
couplings, and finally (4) lead to an observed cross section all mixed
together.  This is alarmingly indirect.  I do not believe that GPD's
and many of the concepts used in GPD's, are physically observable.

\begin{figure}
\epsfxsize=3in\epsfysize=2.4in
\epsfbox{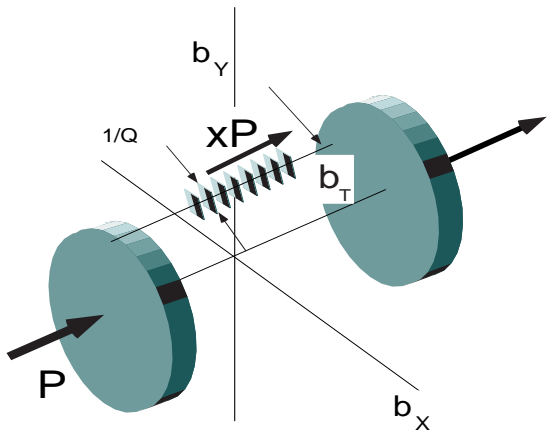}

\caption{Cartoon of the natural interpretation of $GPD$.  The fast
struck quark (stack of plane waves) is located at the transverse
offset position $\bar b_{T}$, while being spatially localized to within
order $1/Q$ in the transverse directions.  The quark's longitudinal
coordinate is not so well localized, but is spread along the
light-cone in the direction conjugate to $xP$.}
\label{fig:tranny}
\end{figure}

If we think about the reaction physically, the situation is remarkable
\cite{femto}.  The electron beam goes right through the proton (Fig
\ref{fig:tranny}) and radiates the detected photon almost instantly.
First, the GPD is not observable, while the DVCS {\it amplitude is
physically observable}.  We should base our thinking on what is
observable.  The reason that an amplitude is observable, for once,
is that the Bethe-Heitler interference term is {\it large}.  This is
{\it good}, because amplitudes contain complete physical information.
It is possible to classify all the amplitudes and their scaling
properties on the basis of angular momentum counting, again assuming
that the {\it quark helicity is conserved.} On this basis it was
predicted \cite{DiehlGousset} that the spin and charge azimuthal asymmetries
of DVCS would go like $sin(\phi)$ and $cos(\phi)$.  The general series
contains terms including $sin(\phi), \, sin(2\phi)$ (spin) and,
$cos(\phi), cos(2\phi)$ (charge) and is extremely complicated but we
made a list.  The simple $sin(\phi)$ and $cos(\phi)$ distributions 
were actually
observed \cite{Azimuthal}, with very little room for higher harmonics.
This is a {\it amazingly beautiful confirmation} of leading order pQCD
in the handbag model.  I must assume that scaling in $Q^{2}$ will be
observed: it is important and necessary as a test of leading order
pQCD. Meanwhile the situation seems very much like the observation of
$\sigma_{L}/\sigma_{T}<<1$ in deeply inelastic scattering: the data is
right on the verge of a breakthrough.

Since the amplitude is what is observable, we should think about how
much information a complete amplitude actually has.  In optics, in
radio astronomy, and in holography, the measurement of an amplitude
allows production of {\it images}.  The process of {\it image
formation} may be unfamiliar and so it is sketched as
follows \cite{femto}: Let a source emit frequency $\omega$ with
amplitude $S(\vec x')$.  Propagate the amplitude from the source to an
observation point $\vec x$.  The proper Green function is the Helmholz
kernel, \ba G_{\omega}(\vec x -\vec x') =\frac{1}{4\pi}\frac{e^{i
\omega|\vec x-\vec x'|}}{|\vec x - \vec x'|} .\ea
This happens to be
{\it just the same on-shell photon kernel} used in particle and
nuclear physics, whose 4-dimensional Fourier transform is
$\delta(k^{\mu}k_{\mu}) \theta(k_{0})$ for the forward propagating
``retarded'' potentials moving into the ``out'' state.  So there is
nothing non-relativistic about the calculation.  To find the
scattering amplitude $M(\vec x)$ at an observation point at $\vec x$
far away compared to the source dimensions, make the ``Fraunhofer''
(asymptotic-out state) approximation \ba 4 \pi M(\vec x) \frac{ e^{i
\omega R}}{R} = \int d^{3}x'\, S(\vec x') \frac{e^{i \omega|\vec x-
\vec x'|}}{|\vec x, \, \vec x'|} \sim \frac{ e^{i \omega R}}{R} \int
d^{3}x'\, S(\vec x') e^{-i \vec k \cdot \vec x'} , \label{labamp} \ea
where $\vec k= \omega \hat x$ is the momentum of the outgoing photons.
The factors of $\frac{ e^{i \omega R}}{R} $ are then removed by
definitions in scattering theory.

Eq. \ref{labamp} is the expression for the amplitude measured in the
lab, with all the conventions cleared away.  Let me repeat that this
amplitude is {\it physically observable} because the Bethe-Heitler
interference acts like a known reference beam, as used in holography
or in optics.  The point is that we can now reconstruct the source:
it is the {\it
inverse Fourier transform} of the measured amplitude.  The image is
the square of the real-space amplitude.  Polarization and spin are
important and described in the literature; at the same time, one is
not obliged to separate all amplitudes in making an image.  The
conversion from ``ray basis'' (momentum states) to ``image basis''
(spatial coordinates) was accomplished in early days by an analog
device called a {\it lens}.

So the classification of amplitudes by alphabetical naming
conventions, etc. is about as interesting as if Rembrandt made
spherical harmonics of his rays.  It may look fancy but will be
exceptionally uninformative after all.  If you have an amplitude, you
cannot beat what will be learnt from making the {\it image}.

\subsubsection{What Will These Images Show? }

$\bullet$ Historically we have only had the longitudinal coordinate
$x=k^{+}/P^{+}$.  The interpretation of what $x$ means in space-time
seems to have been lost.  ``If $\frac{z+t}{\sqrt{2}}$ is the
light-cone time, what the heck is $\frac{z-t}{\sqrt{2}}$.?''

The quandry comes from forgetting that $k^{+}/P^{+}$ is
associated with another {\it asymptotic approximation} of an infinite
Lorentz transformation.  When a sufficiently static field
$\phi(z)_{\alpha }$ is boosted, it transforms to \ba \phi_\alpha(z',
t') \ra \phi_\alpha^\prime(z', t') =\Lambda^{\alpha
\beta}\phi_\beta(z, t), \nn \\
z'=\gamma(z-vt).\ea where $\Lambda^{\alpha \beta}$ is the Lorentz
matrix for the appropriate representation.  In the limit of $v\ra 1$
the field is a function of $z-t= x^{-}$.  The Fourier conjugate
variable is $xP^{+}$.  The big scale $P$ is divided away, just as
scaling away $\gamma$ in $\gamma(z-vt)$ makes Lorentz
boosted pancakes all look the same.  So the physical meaning of $x$
scaling is that naive relativity works, field theory does not destroy
it, the fast pancake is very thin, and the $x$ dependence is showing
us the {\it Fourier transform of the longitudinal structure}.

$\bullet$ Historically the parton model only had forward matrix
elements, quark density matrices \cite{RalSop} of the form $<state | \psi(
x^{\mu })\bar \psi(\vec x^{\mu'})|state>$ found in deeply inelastic
scattering (DIS).  As a consequence of translational invariance, these
correlations depend on {\bf{spatial differences}} $x^{\mu }-x^{\mu'}$.
In GPD's the Feynman $x=(x+x')/2$ dependence tells us the {Fourier
transform of the longitudinal structure {\it between} the locations of
quark interactions.  Since the $x$ dependence ranges
from $0<x<1$, the interaction positions are close together inside the
Lorentz pancake proton.  BUT these variables tell us nothing yet about the
{\it overall location} of the interaction: that would be in the sum $x^{\mu
}+x^{\mu'}$.

$\bullet$ Historically the parton distributions nearly scale in
$Q^{2}$ dependence.  The appropriate frame has $\vec Q $ transverse;
the conjugate spatial variable is the transverse separations of the
quarks $(x^{\mu }-x^{\mu'})_{T}$.  Scaling in $Q^{2}$ says that
once the quarks are close together, and close means nearby compared
to the target size, then nothing else changes.   Logarithmic scaling
violations, the physics of a previous century, reminds us that the
definition of the quark changes very slowly with increasing resolution
$Q^{2}$.  It is described by DGLAP. Meanwhile the {\it overall transverse
location} of the partons cannot be measured or conceived in DIS: the
locations are integrated over by the experiment in taking the limit of
momentum transfer $t=0$.

$\bullet$ Due to DIS, people forgot that the partons were originally
inspired by the Weisczacker-Williams procedure, and always from the
start partons had very definite transverse locations.  The transverse
momentum transfer $\Delta_{\perp}=p_{\perp}-p_{\perp}'$ is conjugate
to the {\it average} transverse spatial components\footnote{Different
conventions exist: Soper's ``center of $P^{+}$" is one.} of $\bar
b_{\perp}=(x^{\mu }+x^{\mu'})_{\perp}$.  The mathematics is Lorentz
covariant, yet best interpreted in a frame where the proton is moving
fast. It is important that $\Delta_{\perp}$ not disturb the system too
much, because this will be the key to establishing an {\it undisturbed
quark radius} in the measurement.  This regime coincides with the one
where the target will be scattered elastically due to strong overlap
with the existing wave functions of the quarks. So measurement of the
$\Delta_{\perp}$ dependence in the lab allows the experimenter
to \cite{femto} {\it scan the transverse image of the proton}.

\medskip

$\bullet$ Each kinematic regime has a useful purpose.  Large $t$,
$Q^{2}=t$, is used for quark counting, but not images.  For images we
want the {\it resolution $\Delta x_{T} \sim 1/Q$} to be small compared
to the target image size $\bar b_{T}\sim 1/\Delta_{\perp}$.  This is
just the regime advocated \cite{DiehlGousset} for the handbag approximation on
more conventional grounds.  Numerically we want $Q^{2}>{\rm GeV}^{2}$, 
so
that the scaling regime is safe, and optical resolution is good.  Bins
of $Q^{2}$ large enough to be safe can be integrated over.  We want
bins in $0< |t|< 1/size^{2}$ to scan across the image, where
presumably $1/size^{2} < {\rm GeV}^{2}$.  \medskip

$\bullet$ Finally there is ``skewness'' $\xi$, the difference of the
longitudinal momentum fractions .  There have been many papers
wondering how to interpret skewness.  Skewness is conjugate to the
{\it Lorentz-rescaled average of the longitudinal positions,} and
allows us to ``take picture at different depths'' through the target:
see Ref. \cite{femto} for the math.

As a rule, the optical resolution should be small compared to the
object scale.  Unfortunately the resolution $1/Q$, the longitudinal
location $\sim 1/\xi P$, and the longitudinal separation $1/xP$ are
all comparable.  With good instruments a lot of information is
extracted, but the Lorentz pancake is resolved on about the same scale
as its thickness.  Some objections based on spectator interactions
\cite{BrodskySpectator} may further weaken the amount one should rely
on the longitudinal information.\footnote{I add this after the
meeting, to address comments by Stan Brodsky.} But wait and see.  The
situation, optically speaking, is similar to taking holograms of an $8
\times 11$ color transparency (and no pun intended).  The overall
physical interpretation of skewness information may be disappointing,
and integrating over both $x$ and $\xi$ to improve statistics is
certainly acceptable.

\medskip

$\bullet$ Meanwhile the quark-Compton scattering kernel is pointlike
in the transverse spatial direction: {\it the image in the transverse
plane is well resolved and reliable.} The number of units of
resolution inside the image size determines image quality: if the
proton is really 1 Fm in size, then $Q\sim 2$  GeV ought to give us
10 units of resolution across the diameter, 100 units across the area.
Even images made with some scaling violations, meaning resolution not
well separated from target size, ought to be useful and interesting.
In retrospect Heisenberg was wrong: the resolution of the Heisenberg
microscope is not the wavelength of the detected photon, and the
disturbance of the target is not the momentum of the photon sent in:
it is possible to make images of elementary particles \cite{XLIU}.
{\it So what will the images show?} Since I am obliged to guess, I
believe that the real proton must be smaller than its poorly resolved
renditions deduced historically from electromagnetic form factors at
low $Q$.

\medskip

$\bullet$ Moreover, to the extent that one can reconstruct the
amplitude of virtual meson production, we can probe the flavor of the
struck quark with good reliability: the $\phi$, the $\pi$, the $\rho$
all have their known quark components, allowing us to {\it take
measurements of the transverse flavor micro-structure of hadrons in
three dimensions.} \medskip

{\it Don't forget the weak probe:} There are other ways to get fine
resolution with small $t$.  I want to suggest that the weak form
factors be pursued in this regard, because the scattering is localized
to $1/M_{W}, \, 1/M_{Z}$ even when $Q^{2}< {\rm GeV}^{2}$.  Now if the
interaction is weak, and fast, we can use the formalism and
interpretation of 40 years ago that assumed it was weak, and measure
an undisturbed quark radius.  In the weak case we have a (light-cone)
matrix element that is {\it off-diagonal} in flavor, of the form $\int
dx d^{2}b_{T}\bar \psi_{u}\psi_{d}e^{i b_{T}\cdot Q_{T}},$ and so on.
We probe both the $u$ and $d$ quarks.  One of the axial form factor of
the proton has a scale of about $1.2 \, {\rm GeV}$, 
substantially above the
$F_{1}$ dipole scale.  Meanwhile the $F_{1}^{EM} $ form factor is
supposedly dominated by $u$ quarks (2 $\times charge \, 2/3$).  If we
naively believed the Fourier transform formulas for both and compare,
then it already says that the $d$ quarks are concentrated closely in
the center.  This is really interesting.  (But I do not believe the
low-$Q^{2}$ Fourier-charge density connection.) Interestingly, the
weak case satisfies the conjecture that the undisturbed
proton is small.  How interesting if the ``weak'' proton is small ,
the ``strong'' one is big, and the DVCS proton (weighted by
charges-squared!)  is different again!

\medskip

We are just starting to find out {\it what might be known.}

\section{Concluding Remarks}

Form factors brought us a long way.  Every time an experiment measures
a definite matrix element, it is a silver sword that pins the theory
so it can be falsified.  Hadron helicity flip pins the ASD model, and
falsifies it; more general pQCD survives.  The new data on the purest helicity
flip imaginable, the electromagnetic form factor $F_{2}$, is very
exciting.

  From $F_{2}$ we believe we are seeing quark orbital angular momentum.
The deduction is very general, yet indirect. Large $A$ targets can
test the predictions.

GPD's are unobservable, but the DVCS amplitude is observable.  We
really want the DVCS amplitude more than the GPD because the
amplitude can make an image.  Nothing any more depends on a
prolonged and uncertain process of extracting every independent amplitude, or
relying on chains of flakey theory models to interpret data.  The
experimentalists can be in charge of getting their own amplitudes and
sending back the ``femtophotography'' images of the target
microstructure.  The existing labs are starting to be real particle
microscopes, and we are all cheering for world {\it femtoscope}
facilities of the future.  There is no limit to the target choice: the
deuteron will be fascinating \cite{BergerPire}.

Then what will the proton look like?  The debate over orbital angular
momentum, if not already one-sided, will be conclusively settled by
images with the proton spin transverse, and the proton appearing
oblong.  Oblateness, meaning the absence of rotational symmetry in the
density matrix, is the last word on OAM. No sum rules, which are
unobservable, are needed.  Quark flavor and gluon substructure may
well be localized by experiments.  The quark OAM may even be localized
within certain substructures of the target image.  How big will the
proton be?  When finally well-resolved, I am betting the proton will
be smaller than the strongly polarized deductions based on the form
factor: maybe 1/2, or 1/5 Fm.  These are very challenging times.

\medskip

What could be more exciting?

\medskip

{\bf{ Acknowledgements: }} Work supported in part under Department of
Energy Grant.

\end{document}